# Linear vs. Nonlinear Diffusion
# and
# Martingale Option Pricing

J.L. McCauley[+], G.H. Gunaratne[++], and K.E. Bassler

Physics Department
University of Houston
Houston, Tx. 77204
jmccauley@uh.edu

[+]Senior Fellow
**COBERA**
Department of Economics
*J.E.Cairnes Graduate School of Business and Public Policy*
NUI Galway, Ireland

[++]Institute of Fundamental Studies
Kandy, Sri Lanka

## Abstract

First, classes of Markov processes that scale exactly with a Hurst exponent H are derived in closed form. A special case of one class is the Tsallis density, advertised elsewhere as 'nonlinear diffusion' or 'diffusion with nonlinear feedback'. But the Tsallis model is only one of a very large class of linear diffusion with a student-t like density. Second, we show by stochastic calculus that our generalization of the Black-Scholes partial differential equation (pde) for variable diffusion coefficients is equivalent to a Martingale in the risk neutral discounted stock price. Previously, this was proven for the restricted case of Gaussian logarithmic returns by Harrison and Kreps, but we prove it here for large classes of empirically useful and theoretically interesting returns models where the diffusion coefficient D(x,t) depends on both logarithmic returns x and time t. Finally, we prove that option prices blow up if fat tails in returns x are included in the market distribution.

# 1. Markov processes with scaling densities

## 1.1 Hurst exponents

Hurst exponents are normally asumed to imply auto-correlations that describe long term memory. We will explain why $H \neq 1/2$ is consistent both with Markov processes, which by construction have no autocorrelations, and also with fractional Brownian motion, where the autocorrelations extend over arbitrarily long nonoverlapping time intervals. Therefore, the Hurst exponent taken alone cannot be used to determine the existence of long term memory.

We begin with the fact that a Hurst exponent H describes, without other assumptions [1], the scaling of the variance of a stochastic process x(t),

$$\sigma^2 = \int_{-\infty}^{\infty} x^2 f(x,t) dx = ct^{2H} \quad (1)$$

where c is a constant. Here, the initial point in the time series is assumed to be $x_o=0$ at t=0. Initially, we limit our analysis to a drift-free process so that $<x>=0$.

To obtain variance scaling (1) we need a probability density f(x,t) that scales as [2]

$$f(x,t) = t^{-H} F(u); u = x/t^H. \quad (2)$$

Up to this point we have not assumed any underlying stochastic dynamics. Our discussion so far is consistent with both fractional Brownian motion (fBm) [1] and a Markov process [2].

## 1.2 Markov processes with scaling

A Markov process [3,4] is a stochastic process without memory, without autocorrelations in x(t). Any stochastic differential equation ('Langevin equation')

$$dx = R(x,t) + \sqrt{D(x,t)} dB(t) \quad (3)$$

with functions R(x,t) and D(x,t) (as opposed to functionals) as drift and diffusion coefficients generates a Markov process [2,5,6]. For the time being we take R=0. B(t) is the usual Wiener process, and Ito calculus [5] is assumed.

From the sde (3) we can also calculate the variance as

$$\sigma^2 = \int_0^t ds \int_{-\infty}^{\infty} dx f(x,s) D(x,s), \quad (1b)$$

so that diffusion coefficient scaling

$$D(x,t) = t^{2H-1} D(u), u = x/t^H \quad (4)$$

follows from combining (1) and (2) with (1b).

The sde (3) with R=0 generates the Green function g of the Fokker-Planck pde [2,7]

$$\frac{\partial f}{\partial t} = \frac{1}{2} \frac{\partial^2}{\partial x^2}(Df), \quad (5)$$

which can be rewritten to include a constant drift rate R by replacing x by x-Rt [8]. Drifts of the form $R(x,t)=\mu-D(x,t)/2$ do not scale, and arbitrary choices of R(x,t) also do not scale. Scale invariance is generally broken by a drift rate depending on x.

We restrict ourselves here to the case where $g(x,0;t,0)=f(x,t)$ because that only these solutions exhibit exact scaling (1,2,4), and also because the density f(x,t) is what one obtains directly from histograms of finance market returns data [7,9]. The full Green function $g(x,x_o;t,t_o)$ is needed in principle for risk neutral option pricing [7] but doesn't scale [2] and also can't be calculated in closed analytic form when D(x,t) depends on both x and t.

Inserting the scaling forms (2) and (3) into the pde (5), then we obtain [2]

$$2H(uF(u))' + (D(u)F(u))'' = 0 \quad (6)$$

whose solution is given by

$$F(u) = \frac{C}{D(u)} e^{-2H \int u du / D(u)}. \quad (7)$$

This generalizes our earlier results [10] to include $H \neq 1/2$.

Next, we study the class of quadratic diffusion coefficients

$$D(u) = d'(\varepsilon)(1 + \varepsilon u^2). \quad (8)$$

This choice generates the two parameter class of student-t- like densities

$$F(u) = C'(1 + \varepsilon u^2)^{-1 - H/\varepsilon d'(\varepsilon)} \quad (9)$$

with fat tail exponent $\mu = 2 + 2H/\varepsilon d'(\varepsilon)$, where H and ε are independent parameters to be determined empirically. So we now know where student-t-like distributions come from dynamically. With $d'(\varepsilon)=1$ we obtain the generalization of our earlier prediction [10] to arbitrary H, 0<H<1, with $\mu=2+2H/\varepsilon$. In that case, we can generate all fat tail exponents $2<\mu<\infty$ (2<μ<7 is observed [11]), and also obtain a finite variance scaling as $\sigma^2 = ct^{2H}$ whenever μ>3. For large u this model fits the fat tails in financial data for all times t [12]. We can generate both the exponential middle [12] and the fat tailed extremes systematically by using a diffusive model.

**3.3 H≠1/2, taken alone, does not imply aucorrelations**

Consider the Markov process (3) in its integrated form

$$\Delta x(t, t+T) = \int_t^{t+T} \sqrt{D(x(s), s)} dB(s) \quad (10)$$

Even when scaling (1,2,4) holds with $H \neq 1/2$, then there can be no autocorrelation in the increments Δx(t,t-T)=x(t)-x(t-T), Δx(t,t+T)=x(t+T)-x(t) over the two nonoverrlapping time intervals [2]. If the nonstationary stochastic process x(t) has *nonstationary increments*, meaning that x(t+T)-x(t)≠x(T) depends both on T and t, then H≠1/2 does not imply autocorrelations [2]. As always with stochastic equations, equality is understood as 'equality in distribution'. The argument that H≠1/2 implies long time correlations fails for Markov processes (and for the finance market as well, to zeroth order) precisely because the stochastic integral (10) with the scaling forms (1,2,4) describes a nonstationary process with *nonstationary increments* whenever H≠1/2:

$$x(t+T) - x(t) = \int_0^{t+T} \sqrt{D(x(s),s)}\, dB(s) - \int_0^{t} \sqrt{D(x(s),s)}\, dB(s)$$
$$= \int_t^{t+T} \sqrt{D(x(s),s)}\, dB(s) = \int_t^{t+T} |s|^{H-1/2} \sqrt{D(u)}\, dB(s)$$
$$= \int_0^T |s+t|^{H-1/2} \sqrt{D(x/|s+t|^H)}\, dB(s) \neq x(T) \quad (11)$$

Direct calculation of the autocorrelation function [2] shows that it vanishes independently of the value of H. *When the increments are nonstationary, then performing a correct data analysis nontrivial [12].*

To obtain the long time autocorrelations of fractional Brownian motion (fBm) [1] two *independent* conditions must be satisfied: (i) the time series must have stationary increments, $x(t+T)-x(t)=x(T)$, and (ii) the variance must scale, $<x^2(t)>=ct^{2H}$ [2]. In the literature these two separate requirements are often confused together into a soup.

**3.8 The Tsallis Model is not generated by 'nonlinear diffusion'**

Many papers have been written claiming that the Tsallis density is generated by both a 'nonlinear Fokker-Planck pde' and by a Langevin eqn. [13-17] as well, but apparently none of the authors noted the following: a Langevin eqn. with ordinary functions as drift and diffusion coefficients is Markovian, generates a linear diffusion pde, the usual Fokker-Planck pde [2,4,5]. The Green function, or condidional probability density, is required to define a Markov process [4,5,7], but a nonlinear diffusion eqn. has no Green function. So, either the Tsallis model is nonlinear but has no underlying Langevin description, or else the Tsallsi density is a Markov process in a nonlinear disguise. We can clear up the confusion.

Given the Tsallis density [17]

$$f_q(x,t) = (c(2-q)(3-q))^{-H} t^{-H} (1+(q-1)x^2/C^2(q)t^{2H})^{1/(1-q)} \quad (11)$$

with H=1/(3-q), where

$$C(q) = c^{(q-1)/2(3-q)} ((2-q)(3-q))^H \quad (12)$$

with c a constant then the fat tail exponent, $f(x,t) \approx x^{-m}$ for x>>1, is µ=2/(q-1), and H=1/(3-q). The Tsallis model definition [17]

$$D_q(x,t) = f_q^{1-q}(x,t) \quad (14)$$

yields

$$D_q(x,t) = (c(2-q)(3-q))^{2H-1} t^{2H-1} (1+(q-1)x^2/C^2(q)t^{2H}) \quad (15)$$

which we can rewrite as

$$D_q(x,t) = d(q) t^{2H-1} (1+((q-1)/C^2(q))u^2) = t^{2H-1} D_q(u) \quad (16)$$

To compare (11,16) with (9,8), we need only write ε=(q-1)/C²(q) and d'(ε)=d(q). Our Fokker-Planck-generated density f(x,t) given by (8) reduces *exactly* to (14) if H=1/(3-q). This means that the Tsallis density $f_q$ actually derives from the *linear* diffusion pde

$$\frac{\partial f}{\partial t} = \frac{1}{2} \frac{\partial^2 (D_q f)}{\partial x^2} \quad (17)$$

so that the so-called 'nonlinear Fokker-Planck eqn.' [13-17] is really a *linear* pde *disguised* superficially as a nonlinear one.

A similar nonlinear disguise is possible for our entire two-parameter student-t-like class solutions (9), because for quadratic diffusion D(u)=d'(ε)(1+εu²), *the solution of the Fokker-Planck pde (5) is a power of the diffusion coefficient,* $F(u) = CD(u)^{-1-H/\varepsilon d'(\varepsilon)}$. *All of these solutions also trivially satisfy a nonlinear pde*, but rewriting (19) as a nonlinear pde in the case of quadratic diffusion superficially masks the real Markovian nature of the dynamics. Because the Tsallis density (11) describes Markovian dynamics, it cannot describe long-time correlated signals like fBm. Here, $H=1/(3-q)\neq 1/2$ merely signals that the increments x(t) are nonstationary.

Next, we explain why nonlinear diffusion would be inconsistent with risk neutral option pricing: the Green function for the Black-Scholes pde solves our market Fokker-Planck pde (5) with a nonscaling drift term $R(x,t)=\mu-D(x,t)/2$ included [7,9].

**2. Pricing options via Martingales**

**2.2 The Black-Scholes PDE and Kolmogorov's First PDE**

With the transformation

$$u = e^{r(t-T)}v \quad (18)$$

the generalized Black-Scholes pde (see [18] for the original Gaussian returns model) in the returns variable [7,9]

$$ru(x,t) = \frac{\partial u(x,t)}{\partial t} + (r - D(x,t)/2)\frac{\partial u(x,t)}{\partial x} + \frac{D(x,t)}{2}\frac{\partial^2 u(x,t)}{\partial x^2} \quad (19)$$

becomes

$$0 = \frac{\partial v}{\partial t} + (r - D(x,t)/2)\frac{\partial v}{\partial x} + \frac{D(x,t)}{2}\frac{\partial^2 v}{\partial x^2}. \quad (20)$$

This is a very beautiful result. *This pde is exactly the backward time equation, or first Kolmogorov equation [3], corresponding to the Fokker-Planck pde (the second Kolmogorov equation)*

$$\frac{\partial g(x,t;x_o,t_o)}{\partial t} = -((\mu - D(x,t)/2)\frac{\partial g(x,t;x_o,t_o)}{\partial x}) + \frac{D(x,t)}{2}\frac{\partial^2 g(x,t;x_o,t_o)}{\partial x^2}$$
(21)

*for the market Green function of returns g, if we choose μ=r in the latter* [7]. With the choice μ=r then both pdes are solved by the same Green function g, so that no information is provided by solving the option pricing pde (19) that is not already contained in the Green function of the Market Fokker-Planck equation (21). The Fokker-Planck operator cannot be made self-adjoint via boundary conditions, the forward and backward time pdes are adjoints of each other [3].

To be explicit, according to the theory of backward time integration [19,20] we must understand (20) as

$$0 = \frac{\partial v(x_o,t_o)}{\partial t_o} + (r - D(x_o,t_o)/2)\frac{\partial v(x_o,t_o)}{\partial x_o} + \frac{D(x_o,t_o)}{2}\frac{\partial^2 v(x_o,t_o)}{\partial x_o^2} \quad (22)$$

where $v(x_o,t_o)=g(x,t;x_o,t_o)$ solves the Fokker-Planck pde (21) in (x,t). This is like physics: everything can be calculated when one knows the Green function [7,21].

We can, e.g., use the market Green function g to price calls risk neutrally as

$$C(p,K,T-t) = e^{r(t-T)}\int_{-\infty}^{\infty}(p_T - K)\theta(p_T - K)g(x_T,T;x,t)dx_T, \quad (23)$$

where $x_T=\ln p_T/p_c$ and $x=\ln p/p_c$ where p is the price at present time t (T is the expiration time, and K is the strike price) and $p_c$ is the consensus price, or 'value' [10,22].

It's easy to check that the diffusive contribution D(x,t) to the drift R(x,t)=r-D(x,t)/2 breaks Hurst exponent scaling. However, option pricing is not a strong test of the underlying market dynamics, is unfortunately insensitive to whether we keep or ignore the term D(x,t) in the drift: we've priced options in agreement with traders prices by treating R as constant while ignoring fat tails, using only the exponential density of returns [7,9]. Why fat tails are ignored is explained below.

**2.3 Generalized Black-Scholes describes a Martingale**

The call price is calculated from the Green function $v=g^+(x,t;x_T,T)$ of the pde (20), where the dagger denotes the adjoint. The forward time Kolmogorov pde

$$\frac{\partial g}{\partial T} = -\frac{\partial}{\partial x_T}((r - D(x_T,T)/2)g) + \frac{\partial^2}{\partial x_T^2}(\frac{D(x_T,T)}{2}g) \tag{24}$$

has exactly the same Green function $g(x_T,T;x,t)=g^+(x,t;x_T,T)$. The price sde corresponding to this Fokker-Planck pde (dropping subscripts capital T, for convenience) is

$$dp = rpdt + \sqrt{p^2 d(p,t)}dB \tag{25}$$

where $d(p,t)=D(x,t)$ and r is the risk neutral rate of return. Then with $g(x,t;x_o,t_o)=G(y,t;y_o,t_o)$ (since dx=dy) [21] we obtain

$$\frac{\partial G}{\partial t} = -\frac{\partial}{\partial y}(-\frac{E(y,t)}{2}G) + \frac{\partial^2}{\partial y^2}(\frac{E(y,t)}{2}G) \tag{28}$$

with $E(y,t)=D(x,t)$, which corresponds to the sde

$$dy = -E(y,t)dt/2 + \sqrt{E(y,t)}dB(t). \tag{29}$$

From this we obtain the corresponding price sde (with $y=\ln S(t)/S_c$)

$$dS = \sqrt{S^2 e(S,t)}dB(t) \tag{30}$$

with price diffusion coefficient $e(S,t)=E(y,t)=D(x,t)=d(p,t)$. This shows that the risk neutral discounted price $S=pe^{-rt}$ with $S_c=p_c$ is drift free, is a local Martingale. The condition for a global Martingale is equivalent, in adition, to requiring that the variance is finite.

That the Black-Scholes pde is equivalent to a Martingale in the risk neutral discounted stock price was proven abstractly for the case of the Gaussian returns model [23]. Our proof is not restricted to the unphysical assumption that D(x,t) is independent of x.

## 2.4 Fat tails yield infinite option prices

When R=constant then the reference price locates the maximum of f and the minimum of D [10,23]. The same definition applies for R(x,t)≠constant. Consider the price of a call for the case where $p > p_c$:

$$C(p,K,T-t) = e^{r(t-T)} \int_{\ln K/p}^{\infty} (p_T - K)g(x_T,T;x,t)dx_T \quad . \quad (31)$$

Here, p is the known stock price at present time t<T. We know the Green function analytically only for the case where g(x,t;0,0)=f(x,t), where f is the empirically observed distribution. This is equivalent to evaluating C for the consensus price, or 'value' $p=p_c$,

$$C(p_c,K,T-t) = e^{r(t-T)} \int_{\ln K/p_c}^{\infty} (p_T - K)f(x_T,T)dx_T \quad . \quad (32)$$

This is adequate for making our main point: *the empirically observed density has fat tails in logarithmic returns* [11,12], $f(x,t) \approx x^{-m}$ for x>>1, so we get

$$C(p_c,K,T-t) \approx e^{r(t-T)} \int_{\ln K/p_c}^{\infty} pe^x x^{-\mu} dx = \infty \quad . \quad (33)$$

The option price will diverge for all p, not just for $p=p_c$. If one inserts a finite cutoff in (32) then the option price is very sensitive to the cutoff, so that one can then predict essentially *any* option price. The evidence that traders don't take fat tails into account is that the exponential returns model prices options in agreement with traders' prices [7,9].

Many papers have been written purporting to price options using fat tails. Most make a simple mistake: they use fat tails in price [24] or price increments, not returns. But fat tails in price is an exponential distribution in x and this describes small to moderate returns, not fat tails [7]. One paper [17] starts correctly with fat tails in returns, but then obtains a spurious Gaussian convergence factor by treating the stochastic variable $u=x/t^H$ in a time integral as if it would be both deterministic and drift free [21].

**Acknowledgement**

KEB is supported by the NSF through grants #DMR-0406323 and #DMR-0427938, by SI International and the AFRL, and by TcSUH. GHG is supported by the NSF through grant #PHY-0201001 and by TcSUH. JMC thanks Cornelia Küffner for reading the manuscript and suggesting changes that made it and more readable, and thanks Vela Velupillai for helping us to spread our message within the economics community.

**References**

1.B. Mandelbrot & J. W. van Ness, *SIAM Rev*. **10**, 2, 422,1968.

2. K.E. Bassler, G.H. Gunaratne, & J. L. McCauley, *"Hurst Exponents, Markov Processes, and Nonlinear Diffusion Equations,* Physica **A** (2006).

3. R.L. Stratonovich. *Topics in the Theory of Random Noise,* Gordon & Breach: N.Y., tr. R. A. Silverman, 1963.

4. N. Wax. *Selected Papers on Noise and Stochastic Processes.* Dover: N.Y., 1954.

5. L. Arnold, *Stochastic Differential Equations*. Krieger, Malabar, Fla., 1992.

6. R. Durrett, *Brownian Motion and Martingales in Analysis*, Wadsworth, Belmont, 1984.

7. J.L. McCauley, *Dynamics of Markets: Econophysics and Finance*, Cambridge, Cambridge, 2004.

8. G.H. Gunaratne & J.L. McCauley. *Proc. of SPIE conf. on Noise & Fluctuations 2005,* 5848,131, 2005.

9. J.L. McCauley & G.H. Gunaratne, *Physica* **A329**, 178 (2003).

10. A. L. Alejandro-Quinones, K.E. Bassler, M. Field, J.L. McCauley, M. Nicol, I. Timofeyev, A. Török, and G.H. Gunaratne, *Physica **363A***, 383, 2006.

11. M.M. Dacoroga, G. Ramazan, U.A. Müller, R.B. Olsen, and O.V. Picte, *An Intro. to High Frequency Finance,* Academic Pr., N.Y., 2001.

12. A. L. Alejandro-Quinones, K.E. Bassler, J.L. McCauley, and G.H. Gunaratne, in preparation, 2006.

13. L. Borland, Phys. Rev. E57, 6634, 1998.

14. G. Kaniadakis & P. Quarati, Physica A237, 229, 1997.

15. G. Kaniadakis, Physica A 296, 405, 2001.

16. T. D. Frank, Physica A 331, 391, 2004.

17. L. Borland, Quantitative Finance 2, 415, 2002.

18. F. Black and M. Scholes, *J. Political Economy* 81, 637, 1973.

19. I. Sneddon, Sneddon, *Elements of Partial Differential Equations*, McGraw-Hill, N.Y., 1957.

20. B. V. Gnedenko, *The Theory of Probability*, tr. by B.D. Seckler, Chelsea, N.Y., 1967.

21. J. L. McCauley, G.H. Gunaratne, & K.E. Bassler, submitted, 2006.

22. J. L. McCauley, G.H. Gunaratne, & K.E. Bassler, in *Dynamics of Complex Interconnected Systems, Networks and Bioprocesses*, ed. A.T. Skjeltorp & A. Belyushkin., Springer, NY, 2006.

23. M. Harrison & D.J. Kreps, *Economic Theory* **20**, 381, 1979.

24. S. Markose and A. Alenthorn, *The Generalized Extreme Value (GEV) Distribution, Implied tail Index and Option Pricing*, preprint, 2005.